\documentclass[prx,twocolumn,aps,showpacs,amsmath,amssymb]{revtex4-1}

\usepackage{graphicx}
\usepackage{epstopdf}
\usepackage{dcolumn}
\usepackage{bm}
\usepackage{amsmath}
 
\begin{document}

\title{Ultrafast Terahertz Conductivity Probes of Topologically Enhanced Surface Transport Driven by Mid-Infrared Laser Pulses in Bi$_2$Se$_3$} 

\author
{L. Luo$^{1}$, X. Yang$^{1}$,  X. Liu$^{2}$, Z. Liu$^{1}$, C. Vaswani$^{1}$, D. Cheng$^{1}$, M. Mootz$^{3}$, I.~E.~Perakis$^{3}$, M. Dobrowolska$^{2}$, J. K. Furdyna$^{2}$, and J. Wang$^{1\ast}$}

\affiliation{$^1$Department of Physics and Astronomy and Ames Laboratory-U.S. DOE, Iowa State University, Ames, Iowa 50011, USA. 
	\\$^2$Department of Physics, University of Notre Dame, Notre Dame, IN 46556, USA.
	\\$^3$Department of Physics, University of Alabama at Birmingham, Birmingham, AL 35294-1170, USA.}

\date{\today}

\begin{abstract}
The recent discovery of topology-protected charge transport of ultimate thinness on surfaces of three-dimensional topological insulators (TIs) are breaking new ground in fundamental quantum science and transformative technology. Yet a challenge remains on how to isolate and disentangle helical spin transport on the surface from bulk conduction. Here we show that selective mid-infrared femtosecond photoexcitation of exclusive {\em intraband} electronic transitions at low temperature underpins topological enhancement of terahertz (THz) surface transport in doped Bi$_2$Se$_3$, with no complication from interband excitations or need for controlled doping. The unique, hot electron state is characterized by conserved populations of surface/bulk bands and by frequency-dependent hot carrier cooling times that directly distinguish the faster surface channel than the bulk. We determine the topological enhancement ratio between bulk and surface scattering rates, i.e., $\gamma_\text{BS}/\gamma_\text{SS}\sim$3.80 in equilibrium. These behaviors are absent at elevated lattice temperatures and for high pump-photon frequencies and fluences. The selective, mid-infrared-induced THz conductivity provides a new paradigm to characterize TIs and may apply to emerging topological semimetals in order to separate the transport connected with the Weyl nodes from other bulk bands.
\end{abstract}

\maketitle
\preprint{Manuscript to Nat. Commun.}

\section*{Introduction}

The recent discovery of topology-protected charge transport of 3D TIs has led to a promising platform for exploring both fundamental topological quantum phenomena and technological applications \cite{DHsieh,HZhang,JEMoore,MZHasan,XLQi,NKoirala,PBowlan}. The emergent behaviors associated with this exotic state of matter, such as spin-locking, helical spin structure, topological invariant, chiral anomaly and dissipationless currents \cite{SBravyi,DHsieh2,CWu,CLKane,JEMoore2,AAZyuzin}, offer new perspectives for achieving transformational technological applications in spintronics and quantum sensing, computing and communications \cite{JEMoore} beyond the current technological limit. One of the current frontiers for characterizing and understanding topological phenomena lies in the fundamental challenge of how to isolate and disentangle symmetry-protected transport from bulk conduction.
On the one hand, development and optimization of hot electron transistors and modulators operating at ultra-high THz frequencies will significantly benefit from the direct probing of ultrafast THz charge transport on the surface and from a better understanding of the fundamental effects of topological spin-locking on carrier scattering. 
Although the existence of a helical Dirac spectrum has been well established in electrical transport \cite{ONG} and photoemission \cite{SHEN,DHsieh2,YHWang,DHsieh3}, such experiments are not direct for accessing high-frequency transport and, thereby, ultrafast helical spin transport in the THz spectral region has not been clearly identified yet.  
This is due, in part, to the limitations of conventional transport characterization methods that 
are incapable of probing frequency-dependent conductivity characterized by femtosecond (fs) in time and THz in energy scales.
On the other hand, 
in 3D TIs, the coexistence and mutual scattering of Dirac and bulk carriers lead to an intertwined response. 
Despite recent observations from static THz measurements of TI samples with significantly reduced doping and defects 
\cite{staticTHz, axion}, underpinning the intrinsic surface scattering processes from hot Dirac carriers and their topological enhancement is still challenging, especially in samples that commonly exhibit unintentional high density doping into bulk states.  

The transport properties of 3D TIs are determined by carriers in narrowly gapped bulk bands and by gapless surface states protected by the time-reversal symmetry.  
Although the spin-momentum locking of Dirac electrons increases the conductivity from ``atomically-thin" surface states, its contribution can still be masked by the high density carrier conduction, at either bulk or interface states, in static transport measurements, even in films with 10s of nm thickness. 
Consequently, very few experimental techniques are capable of isolating and studying the intrinsic topological transport on the surface, especially for high Fermi energies. 
Sophisticated characterization and sample synthesis are involved, e.g., anisotropic magneto-transport by adjusting magnetic field directions \cite{ONG} and significantly reduced doping and defects by thin film engineering \cite{axion, YZhang,JLinder,CXLiu,HZLu}. 
All of these methods have either complications with limited applicability or ambiguity in their interpretation. For example, intrinsic Dirac-cones could be contaminated with possible gap opening in very thin samples \cite{YZhang,JLinder,CXLiu,HZLu}. 
The separation of surface and bulk/interface contributions to the static THz conductivity spectra often relies on data fitting with many parameters without sufficient constraints. For example, in many cases, they can be fitted equally well with both single and multiple Drude components \cite{LWu,RVAguilar}. In special cases, a two-dimensional (2D) electron gas at interface can completely dominate the surface transport that may lead to thickness-independent conductivity spectra \cite{RVAguilar}. 

Recently, ultrafast THz conductivity has been shown to be promising by providing time resolution able to distinguish the surface from the bulk responses after suddenly driving the system out-of-equilibrium \cite{RVAguilar,SSim}. 
However, the high photon frequency pumping at 1.55 eV used so far introduces interband excitations from valence to conduction and other high lying bands at both surface and bulk. Such coupling causes population transfer between surface and bulk bands and enhanced scattering between high-energy states.  
These complications have hindered clear observation of intrinsic carrier scattering and dynamics on the surface in spite of several very interesting ultrafast optical pumping studies \cite{RVAguilar,SSim}.

Selective mid-infrared (mid-IR) pump-THz probe spectroscopy represents a powerful and versatile tool, never used before, that allows us to measure surface transport in TIs, even for Fermi level $E_\text{F}$ into the bulk bands.  
We emphasize several strategic advantages of this scheme. First, in contrast to interband photoexcitation that couples high energy electronic states between two bands, mid-IR photoexcitation exclusively excites intraband transitions near $E_\text{F}$. Here the photon energy is well below the insulating bandgap and Fermi energy, illustrated in Fig. 1(a), which at low temperatures suppresses phonon-mediated scattering leading to surface-bulk charge transfer \cite{ YHWang}. This allows measurement of intrinsic scattering rates and relaxation pathways for fixed number of surface and bulk carriers and strongly constrains the fitting  parameters. In particular, only scattering rates are free parameters due to the conserved surface/bulk populations for mid-IR photoexcitation.  Additionally, our time resolution can separate intrinsic surface from bulk conductivity via their different hot carrier cooling times, i.e., resolve the issue directly in the time domain. 
The extracted spectra fully characterize THz response functions as the complex frequency-dependent conductivity. In this way we can determine the intrinsic surface electron scattering rates and their ultrafast dynamics, complimentary to commonly used transport \cite{ONG} and photoemission \cite{SHEN,DHsieh2,YHWang,DHsieh3} measurements.  
Finally, comparison of the conductivity for selective photoexcitations below and above the interband transition enabled by tuning the pump photon energy allows the separation of carrier dynamics within individual surface/bulk bands from surface-bulk charge transfer. The pump photon energy tuning scheme achieved allows us to control the surface transport optically in a selective way. 
 However, in contrast to the several interband photoexcitation studies performed in TIs \cite{RVAguilar,SSim}, mid-IR pump induced conductivity experiments have not been carried out so far to reveal the elusive, ultrafast surface transport discussed above.

\begin{figure*}[!tbp]
\includegraphics[scale=0.6]{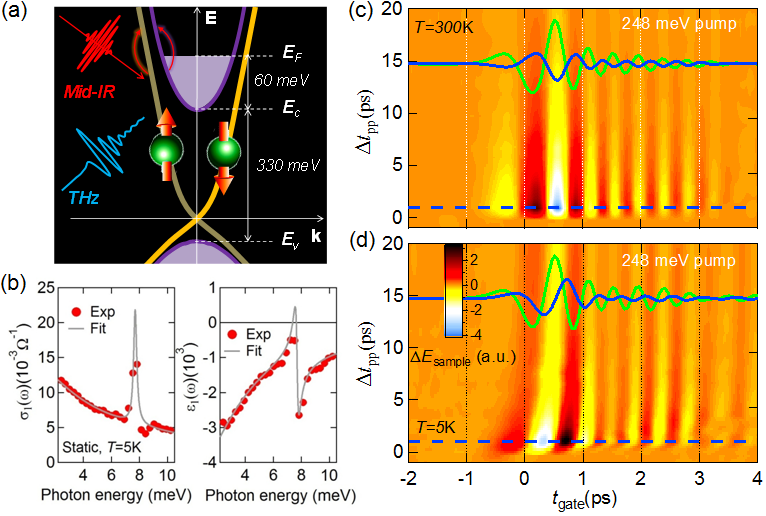}
\caption{\textbf{Ultrafast mid-IR pump and THz probe spectroscopy for Bi$_2$Se$_3$.} (a) Schematic of the surface and bulk electronic band structures of the Bi$_2$Se$_3$ film with Fermi level ($E_\text{F}$) indicated, illustrating mid-IR intraband excitations and THz conductivity detection. (b) Static THz spectra of $\sigma_{1}(\omega)$ and $\varepsilon_{1}(\omega)$ (red dots) of the Bi$_2$Se$_3$ film at $T$=5 K. Shown together are the theoretical fitting (gray lines) using the composite THz model from Eq. ($\ref{equ1}$). (c,d) A 2D false-color plot of pump-induced THz field changes $\Delta E_\text{sample}$ as a function of gate time $t_\text{gate}$ (horizontal axis) and pump-probe delay $\Delta t_\text{pp}$ (vertical axis) after 248 meV (5 $\mu$m) photoexcitation of the sample at (c) $T$=300 K and (d) $T$=5 K, respectively. Plotted together are the corresponding static THz fields $E_\text{sample}(t_\text{gate})$ (green curves) and their pump-induced changes $\Delta E_\text{sample}(t_\text{gate})$ at $\Delta t_\text{pp}$=1 ps (blue curves, $\times$5, from the cut positions as marked by the blue dashed lines), in order to compare their relative amplitude changes and phase shifts at different temperatures.}
\label{Fig1}
\end{figure*}

Here we report ultrafast THz charge transport and dynamics in n-type Bi$_2$Se$_3$ that arise from non-equilibrium, yet conserved electron populations, Dirac surface and bulk transient states induced by intraband fs mid-IR photoexcitation. 
We observe frequency-dependent carrier cooling times of photoinduced THz conductivity that clearly differentiate surface from bulk contributions and allow determination of their scattering rates.
Most intriguingly, we show that the topological enhancement of surface transport suppresses the surface electron scattering rate as compared to the bulk, i.e., $\gamma_\text{BS}/\gamma_\text{SS}\sim$3.80 in equilibrium. 
This result is consistent with surface helical spin transport in the presence of short-range disorder \cite{TOzturk}.  
The distinct spectral-temporal characteristics observed here are absent for elevated lattice temperature, high pump photon frequency and high fluence. Under the latter conditions, the ultrafast conductivity dynamics is characterized by frequency-{\em independent} carrier cooling times due to bulk-surface charge transfer within $\sim$10 ps, which is driven by phonon-mediated scattering due to interband optical excitation and lattice heating. 
We achieve all these results this via selective mid-IR pumping {em only}, i.e. by suppressing all interband transitions that mask the processes of main interest. This unique approach is distinctly different from any prior THz transport measurements so far, which have used either high energy optical frequency photoexcitation or only detected time-averaged properties. 
Separation of the {\em intraband}, mid-IR-pump-induced THz conductivity provides a powerful method for isolating the surface transport channel in TIs -- unhindered by complications from interband pumping or high density doping into bulk states. The distinct spectral-temporal characteristics obtained in this way in TIs may be extended to study and understand much broader topological phenomena, e.g., topological semimetals which is limited by the inability to separate the transport behaviors connected with the Weyl nodes from other (trivial) bulk bands \cite{BYan}.

\section*{Results}
Our mid-IR pump and THz probe spectrometer is driven by a 1 kHz Ti:sapphire regenerative amplifier which has 800 nm central wavelength and 40 fs pulse duration \cite{Luo1}. The majority of the output is used to pump an optical parametric amplifier to generate mid-IR pulses tunable from 3-15 $\mu$m (or 83-413 meV) allowing for selective intra- and inter-band photoexcitation. 
The other part of the output is used to generate and detect phase-locked THz electric fields in time-domain via optical rectification and electro-optic sampling in a 1 mm thick ZnTe crystal, respectively. THz fields with a bandwidth from 0.5THz to 2.5 THz (2-10.3 meV) are used as a probe beam. 
The measurement scheme is briefly illustrated in Fig. 1(a), in which mid-IR pump (red) photoexcites the sample and THz pulse (light blue) probes photoinduced responses of the sample. The transmitted THz pulse containing spectral amplitude and phase information of the sample is directly measured in time-domain using an optical gate pulse.
The Bi$_{2}$Se$_{3}$ thin film sample, 50 nm thick, is grown by molecular beam epitaxy on a 0.5 mm thick sapphire substrate. 
The sample is mounted together with a 0.5 mm thick pure sapphire substrate, used as a reference, into a cryostat with temperatures down to $T$=5 K. 
Two copper mounts with the same aperture are placed directly in front of the sample and reference to ensure uniform photoexcitation and accurate comparison of their THz transmission. The setup is enclosed in a N$_{2}$ gas purge box.
The sample studied has a Fermi energy $E_\text{F}\sim$60 meV from the bulk conduction band edge (Fig. 1(a)), as estimated from the measured plasma frequency in the static THz conductivity spectra shown in Fig. 1(b).
The gray lines are theoretical fittings that are discussed later.
The estimated Fermi energy is consistent with the difference between the separately measured, interband optical transition onset $E_\text{inter}\sim$390 meV using Fourier transform infrared (FTIR) spectroscopy (see Supplementary Material) and the bulk insulating bandgap $E_\text{g}\sim$332 meV \cite{IANechaev}, which yields a similar $E_\text{F}=E_\text{inter}-E_\text{g}\sim$58 meV.  

\begin{figure*}[tbp]
	\includegraphics[scale=0.6]{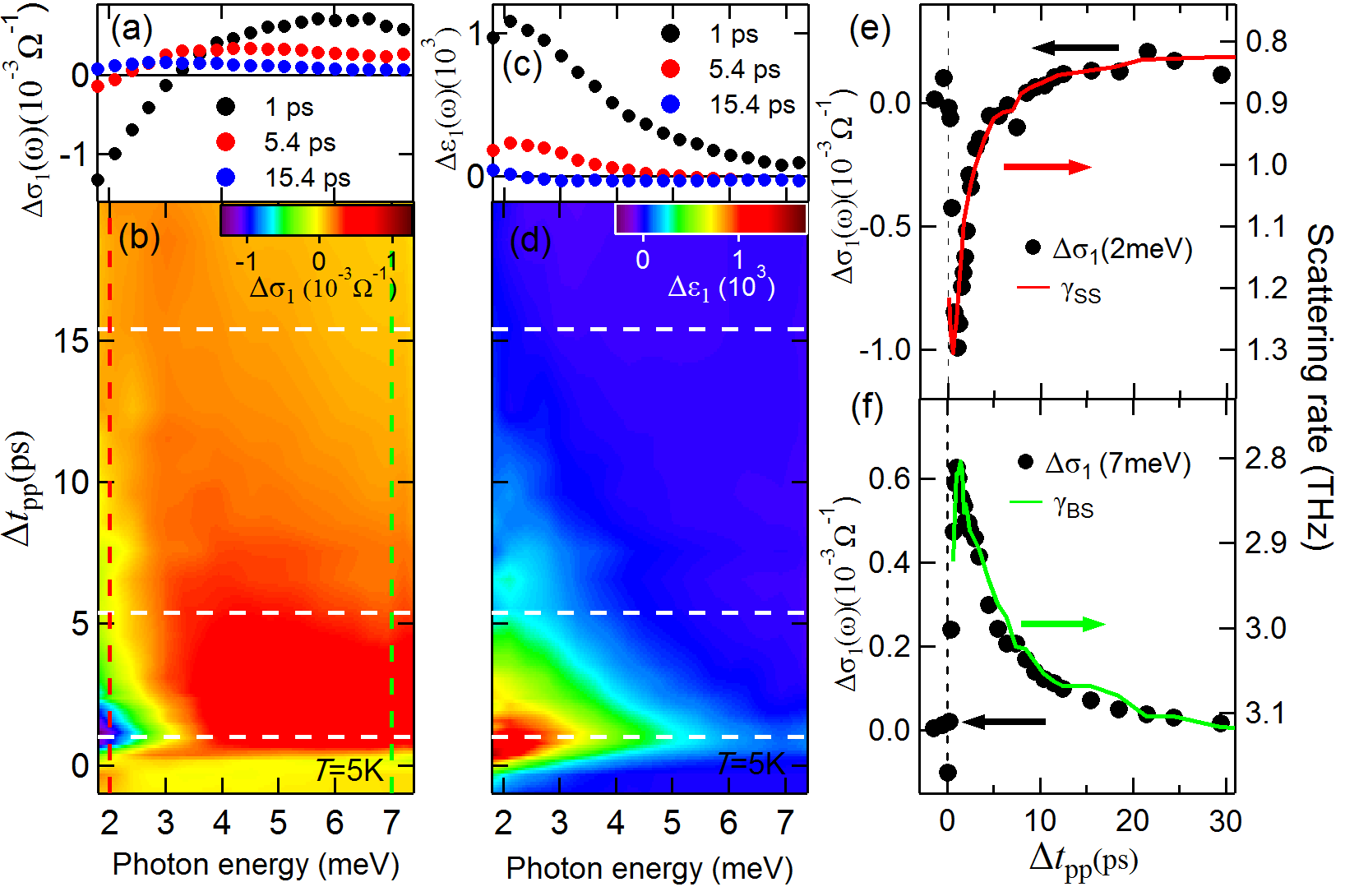}
	\caption{\textbf{Frequency dependent dynamics of THz spectra at low temperature}. Pump-induced THz spectra of (a,b) $\Delta\sigma_1(\omega)$ and (c,d) $\Delta\varepsilon_1(\omega)$ after 248 meV (5 $\mu$m) photoexcitation with fluence 12 $\mu$J/cm$^{2}$ at $T$=5 K as a function of pump-probe delay  $\Delta t_\text{pp}$. (a) and (c) show the THz spectra from three  cut positions from the corresponding 2D plots of (b) and (d), respectively, at $\Delta t_\text{pp}$=1, 5.4, 15.4 ps, as indicated by the white dashed lines. (e) THz conductivity at 2 meV, i.e., $\Delta \sigma_{1}$(2meV) (black dots, left axis), as a function of $\Delta t_\text{pp}$, from the frequency-cut position in (b) as indicated by the red dashed line. The scattering rate of the surface state $\gamma_\text{SS}$ (red curve, right axis) obtained from theoretical fitting using Eq. $\ref{equ1}$ is plotted together to compare relaxation dynamics. (f) Similar to (e), THz conductivity at 7 meV $\Delta \sigma_{1}$(7meV) (black dots, left axis) and the scattering rate of bulk state $\gamma_\text{BS}$ (green curve, right axis) are plotted. $\Delta \sigma_{1}$(7meV) is from the frequency-cut position in (b) as indicated by the green dashed line. As shown clearly, the relaxation dynamics of the surface (bulk) scattering rate matches with that of THz conductivity at 2 meV (7 meV) very well. This frequency dependent THz relaxation dynamics indicates surface (bulk) state is more sensitive to low (high) THz frequency.}
	\label{Fig2}
\end{figure*}  


We characterize the nonequilibrium THz responses of the Bi$_{2}$Se$_{3}$ thin film by extracting the real parts of the transient conductivity $\sigma_{1}(\omega, \Delta t_\text{pp})$ and dielectric function $\varepsilon _{1} (\omega, \Delta t_\text{pp})$ as a function of both the frequency $\omega$ and the pump-probe delay time $\Delta t_\text{pp}$. 
These spectra describe the dissipative and inductive parts of the response functions of quasi-particles, respectively, and are extracted from raw THz fields in time-domain, which are measured as a function of gate time $t_\text{gate}$ after transmission through (i) the reference bare substrate $E_\text{ref}(t_\text{gate})$, (ii) the unexcited sample $E_\text{sample}(t_\text{gate})$, and (iii) as pump-induced change $\Delta E_\text{sample}(t_\text{gate},\Delta t_\text{pp})$ after pump-probe delay $\Delta t_\text{pp}$.
Figs. 1(c)-1(d) show the 2D false-color plots of $\Delta E_\text{sample}(t_\text{gate},\Delta t_\text{pp})$ after below gap  photoexcitation at 248 meV (5 $\mu$m) for high (300 K) and low (5 K) temperatures, respectively. 
Also plotted in the above figures are the corresponding $E_\text{sample}(t_\text{gate})$ (green curves) and pump-induced change $\Delta E_\text{sample}(t_\text{gate})$ at $\Delta t_\text{pp}$=1 ps (blue curves).
For the unexcited sample, through the fast Fourier transformation and Fresnel equation, the static THz conductivity, shown in Fig. 1(b), is directly obtained from the experimentally measured complex transmission coefficient $\tilde{t}(\omega)$ with spectral amplitude and phase information by comparing the fields transmitted through sample and reference, i.e., $\tilde{t}(\omega)=E_\text{sample}(\omega)/E_\text{ref}(\omega)$ without the need for Kramers-Kronig transformation. 
For the excited sample at $\Delta t_\text{pp}$, the complex transmission coefficient is obtained by $\tilde{t}_\text{excited}(\omega)=[\Delta E_\text{sample}(\omega)+E_\text{sample}(\omega)]/E_\text{ref}(\omega)$ and is used to simultaneously extract the photoexcited transient conductivity and dielectric function. 
Afterwards, the pump-induced change $\Delta\widetilde{\varepsilon}(\omega)=\widetilde{\varepsilon}_\text{excited}(\omega)-\widetilde{\varepsilon}(\omega)$ is extracted. 
The corresponding complex conductivity can be calculated by the equation $\widetilde{\sigma}(\omega)=i[1-\widetilde{\varepsilon}(\omega)]\omega{\varepsilon}_{0}$. 
The simultaneously obtained real parts of transient conductivity change  $\Delta\sigma_{1}(\omega)$ and dielectric function change  $\Delta\varepsilon_{1}(\omega)$ as a function of $\Delta t_\text{pp}$ are presented in the study, which allow us to quantitatively describe the ultrafast dynamic evolution of the photoexcited Dirac and bulk fermions and their intrinsic scattering processes.  

The pump-induced raw THz fields in Figs. 1(c) and 1(d) exhibit a distinctly different temperature dependence. 
At 300 K, the transmitted field change $\Delta E_\text{sample}^{}$ has a $\pi$ phase shift relatively to the static field $E_\text{sample}^{}$ without any other frequency- and time-dependent changes. This behavior is characteristic of an absorption-dominated response, i.e., a large increase in $\sigma_1(\omega)$. 
On the other hand, at low temperature $T$=5 K, a large inductive response $\varepsilon_1(\omega)$ appears in addition to the dissipative one, as manifested by a significant frequency- and time-dependent THz field reshaping besides the dominant $\pi$ phase shift as in the 300 K case. These distinct spectral-temporal characteristics give rise to several salient features in the extracted, THz response functions $\Delta\sigma_1(\omega)$ and $\Delta \varepsilon_1(\omega)$, shown in Figs. 2(a)-2(d), that allow us to separate and measure the intrinsic surface and bulk transport contributions to the conductivity without reference to any theoretical model as discussed next.   

\begin{figure}[h]
	\includegraphics[width=86mm]{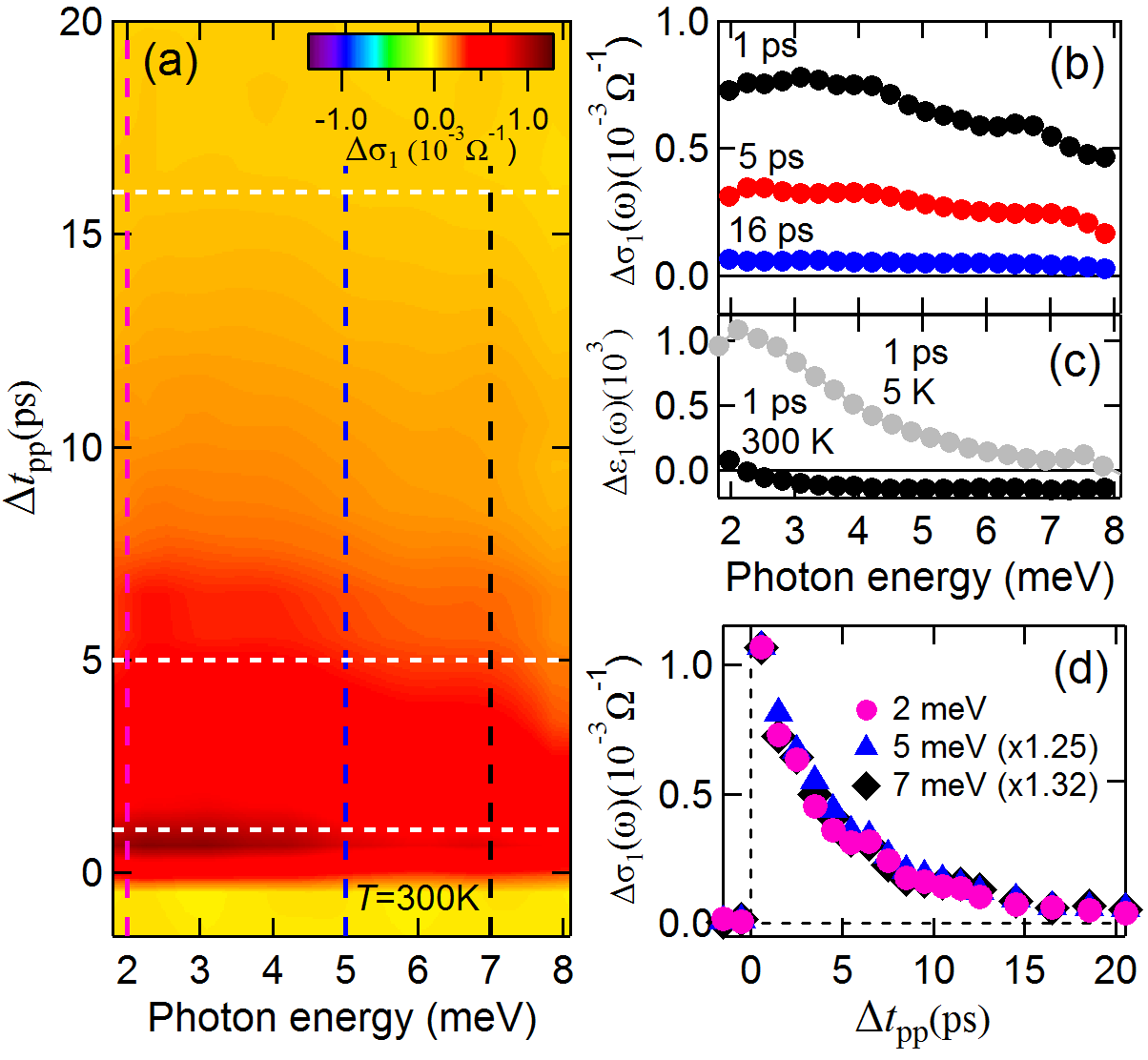}
	\caption{\textbf{Frequency independent dynamics of THz spectra at high temperature.} (a) Pump-induced THz spectra of $\Delta\sigma_1(\omega)$ after 248 meV (5 $\mu$m) photoexcitation with fluence 12 $\mu$J/cm$^{2}$ at $T$=300 K as a function of pump-probe delay $\Delta t_\text{pp}$. (b) THz conductivity $\Delta\sigma_1(\omega)$ from three cut positions from (a), at $\Delta t_\text{pp}$= 1, 5, 16 ps, as indicated by the white dashed lines. 
		(c) THz spectra of $\Delta\varepsilon_1(\omega)$ at $\Delta t_\text{pp}$=1 ps for 5 K (gray dots) and 300 K (black dots), the amplitude of which is much smaller at high temperature, opposite to low temperature ones in Figs. 2(c) and 2(d). (d) The comparison of THz conductivity $\Delta\sigma_1(\omega)$ at three frequencies: 2, 5, 7 meV, as a function of $\Delta t_\text{pp}$ (scaled to match amplitude), from three frequency-cut positions from (a) as indicated by magenta, blue, and black dashed lines, respectively. The frequency independent relaxation dynamics of $\Delta\sigma_1(\omega)$ indicates the strong coupling of surface and bulk states that make them decay together at high temperature.  
	}
	\label{Fig3}
\end{figure}

The first feature, shown in Figs. 2(a)-2(b), is that the pump-induced THz conductivity spectra $\Delta\sigma_1(\omega)$  extracted from the measured time-dependent THz fields exhibit a clear bi-polar behavior with frequency-dependent cooling times under {\em intraband} excitation (248 meV) at low temperature (5 K). 
Specifically, as shown by comparing three time-cut spectra (white dashed lines in Fig. 2(b)), $\Delta\sigma_1(\omega)$ at early times ($\Delta t_\text{pp}\textless$5 ps) is characterized by a strong bleaching, i.e., a negative conductivity change, in the low frequency 2-3 meV range. This changes to a strong absorption, i.e., positive conductivity change, towards high frequencies up to $\sim$7 meV, as seen in the 1 ps trace (black dots in Fig 2(a)). At $\Delta t_\text{pp}\textgreater$5 ps, the low frequency bleaching has quickly diminished, and the transient spectra are dominated by the high frequency absorption with longer cooling times, e.g., as shown in the 5.4 ps (red dots) and 15.4 ps (blue dots) traces. The frequency cuts of the $\Delta\sigma_1(\omega)$ spectra from Fig. 2(b) are summarized in Figs. 2(e) (at $\omega$=2 meV) and 2(f) (at $\omega$=7 meV) (black dots) as a function of $\Delta t_\text{pp}$. Clearly, the cooling times depend on the probe frequency, i.e., the faster (slower) cooling for ultrafast THz conductivity change at low (high) frequency.  
The characteristic cooling time for the 2 meV probe is $\tau_\text{SS}$=1.48 $\pm$ 0.14 ps, significantly shorter than that for the 7 meV probe $\tau_\text{BS}$=5.30 $\pm$ 0.18 ps. 
Such a frequency dependent decay of THz electron transport is consistent with the significant THz pulse reshaping seen in Fig. 1(d), which indicates more than one cooling channel of hot electrons. 

Second, to understand these experimental results, we recall that prior ultrafast THz responses from photoexcited graphene with similar Dirac conical band structure have been shown to exhibit a similar transient bleaching of THz conductivity \cite{GJnawali}. 
While much more complicated dynamics are reported in Bi$_{2}$Se$_{3}$ samples, where a decrease in their thickness has been shown to induce a transition to a ``surface-like", pronounced bleaching behavior that differs from a ``bulk-like", absorption one \cite{SSim}.
The latter is expected from the ionized impurity scattering in the bulk which decreases scattering rate after pumping at low temperatures 
\cite{DChattopadhyay} and gives rise to a positive $\Delta\sigma_1(\omega)$ towards zero frequency (refer to Fig. 4 for details). 
Therefore our results show that THz transport and dynamics of surface and bulk fermions can be selectively separated in time and measured by tuning probe frequency. 
Most intriguingly, upon closer scrutiny, the time-dependent scattering rates of surface Dirac (red line) and bulk (green line) fermions in Figs. 2(e) and 2(f) match very well with the conductivity dynamics at 2 meV and 7 meV, respectively. These scattering rates are extracted by fitting the time-dependent conductivity data as discussed later.  
This clearly shows that the experimentally obtained $\Delta\sigma_1(\omega)$ probed at 2 meV can be used to directly measure surface transport dynamics, distinctly different from bulk conduction (at 7 meV), without reference to theoretical models. 

The third feature is that increasing the lattice temperature leads to an induced absorption behavior over the entire frequency range measured with a frequency-{\em independent} single cooling time, as shown in Fig. 3. Strikingly, this high-temperature behavior is opposite to the low temperature conductivity responses in Fig. 2. This is consistent with the different, raw THz fields $\Delta E_\text{sample}^{}$ between $T$=5 K and 300 K seen in Figs. 1(c)-1(d). 
Specifically, the pump-induced $\Delta\sigma_1(\omega)$ spectra at 300 K, shown as 2D false color plot in Fig. 3(a), are all positive, $\Delta\sigma_1(\omega)>$0, without the negative bleaching component over the measured time and spectral range. The relative change of conductivity, $\Delta\sigma_1/\sigma_1$ at 300 K dominates that of dielectric function, $\Delta\varepsilon_1/\varepsilon_1$. For example, $\Delta\sigma_1(\omega)$ has similar amplitude at 5 K and 300 K as shown in Figs 2(a) and 3(b), but $\Delta\varepsilon_1(\omega)$ at 300 K is decreased to $\sim$one order of magnitude smaller than its counterpart at 5 K, as shown in Figs. 2(c) and 3(c) (also see the Supplementary Material). Since phonon-mediated coupling and charge transfer between surface and bulk are activated above the Debye temperature at 182 K \cite{GEShoemake,YHWang}, the dominance of positive $\Delta\sigma_1(\omega)$ at 300 K is consistent with our conclusion that the bulk transport and surface-bulk charge transfer dominate at high temperatures and lead to the induced absorption. This suppresses the induced bleaching from the surface seen at 5K in Fig. 2(a). 
One the other hand, the frequency dependent cooling times previously seen at 5 K are no longer present at the elevated temperature. As shown in Fig. 3(d) the THz conductivities $\Delta\sigma_{1}(\omega)$ at $\omega$=2, 5, and 7 meV as a function of pump-probe delay $\Delta t_\text{pp}$, now decay with the same time at 300 K. The cooling time of $\sim$4.8 ps is similar to the bulk state cooling time $\tau_\text{BS}$ observed at 5 K. 
These observations corroborate again our claim that the low frequency conductivity at 2 meV, with faster cooling time observed at 5 K, comes from surface transport that is separated from the bulk channel.  
Furthermore, these distinct differences between high and low temperatures in ultrafast charge transport suggest that surface-bulk charge transfer in TIs is suppressed at low temperature, i.e., conserved transient populations within surface and bulk bands, because of lacking phonon-mediated scattering process.   

To put these physical pictures on a strong footing, we quantitatively simulate the experimentally determined THz response functions $\Delta\sigma_{1}(\omega)$ and $\Delta\varepsilon _{1} (\omega)$. By extracting the temporal evolution of surface and bulk conductivity spectra, we obtain the intraband scattering rates and ultrafast dynamics of both Dirac (red line) and bulk (green line) fermions, as presented in Figs. 2(e) and 2(f), respectively.
A theoretical model consisting of contributions with Drude and Lorentzian lineshapes is used to reproduce the measured THz response functions. 
Specifically, two Drude terms account for surface (SS) and bulk (BS) states plus one phonon oscillator centered at $\omega_{0}\sim$7.7 meV (1.9 THz),

\begin{eqnarray}  \label{equ1}
\begin{aligned}
&\widetilde{\varepsilon}^{\text{}}(\omega) = \sum_{j=\text{SS,BS}}\widetilde{\varepsilon}^\text{Drude}_{j}(\omega) + \varepsilon_\text{ph}(\omega)= \\
&\varepsilon_{\infty}-\sum_{j=\text{SS,BS}}\frac {(\omega_\text{p})_{j}^{2}} {\omega^{2}+i\omega\gamma_{j}}+ \frac{F}{(\omega^{}_{0})^{2}- \omega^{2}-i\omega\Gamma},
\end{aligned}
\end{eqnarray}
where $\varepsilon_{\infty}$ is the background electrical permittivity and the plasma frequency $(\omega_{\text{p}}^{2})_{j}=ne^{2}/\varepsilon_{0}m^{*}$ is proportional to the density of charge carriers of surface ($n_\text{SS}$, $j$=SS) and bulk ($n_\text{BS}$, $j$=BS) electrons. $\omega$ and $e$  are the frequency and electron charge. $\varepsilon_{0}$ and $m^{*}$ are the vacuum electrical permittivity and carrier effective mass. $\gamma_{j}$ is the scattering rate of surface ($j$=SS) and bulk ($j$=BS) carriers. In the last term of Eq. $\ref{equ1}$, $F$ denotes effective transition strength of the optical phonon with resonance frequency $\omega_{0}$ and scattering rate $\Gamma$. 
Such a composite THz response model provides an excellent agreement with the pump-induced transient THz spectra over the entire measured pump-probe delays and spectral range as shown in Fig. 4 (cyan lines). 
Such fit is achieved by only varying the scattering rates $\gamma$ of surface and bulk carriers, i.e., the charge carrier densities of surface ($n_\text{SS}$) and bulk ($n_\text{BS}$) states remain constant in the fits. 
Please also note that the optical phonon resonance $\sim$7.7 meV is fitted by $F$, $\omega_{0}$, and $\Gamma$ of the Lorentzian oscillator which only affect locally feature near the resonance. 
Examples of typical fits are shown in Fig. 4 at various time delays, $\Delta t_\text{pp}$=1 ps, 5.4 ps, and 15.4 ps. The model fit (cyan lines) is divided into the  sum of the surface (green dashed lines), bulk (blue dashed lines) and phonon (magenta dashed lines) responses. 

\begin{figure}[h]
\includegraphics[width=86mm]{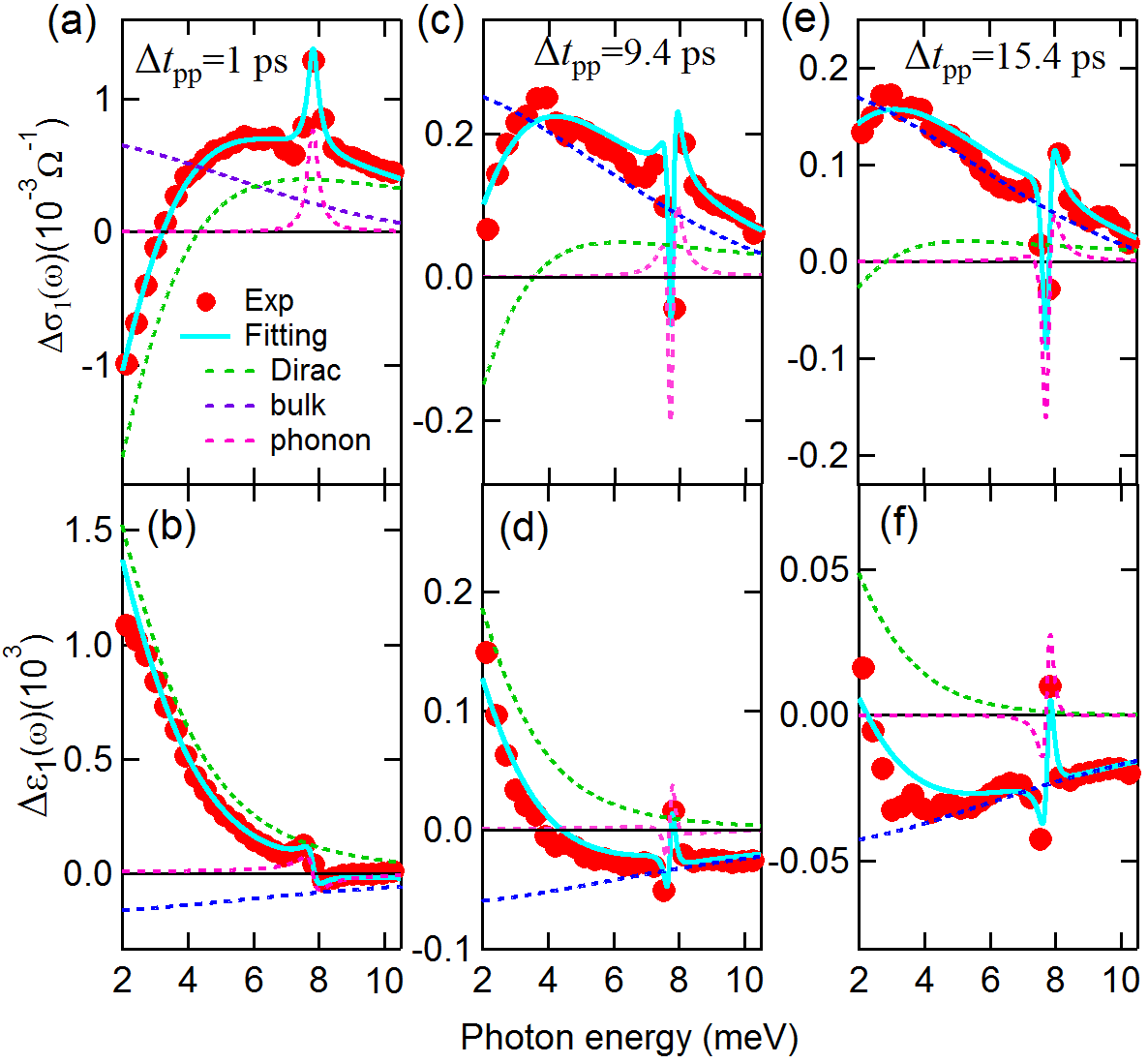}
\caption{\textbf{Theoretical fitting of the pump-induced THz spectra.} (a,b) The fitting of $\Delta\sigma_1(\omega)$ and $\Delta\varepsilon_1(\omega)$ for 248 meV (5 $\mu$m) photoexcitation with fluence 12 $\mu$J/cm$^{2}$ at $T$=5 K and pump-probe delay $\Delta t_\text{pp}$=1 ps. The experimental results (red dots) are fitted very well by the THz model from Eq. $\ref{equ1}$ (cyan lines), which consists of three individual components, i.e., the Dirac surface state (green dashed lines), bulk state (blue dashed lines), and optical phonon mode (pink dashed lines). The fitting results for (c,d) $\Delta t_\text{pp}$=5 ps and (e,f) $\Delta t_\text{pp}$=16 ps are plotted in the same manner as (a,b). 
}
\label{Fig4}
\end{figure}

We like to emphasize three key observations from the quantitative fitting. First, while the static THz spectra, shown in Fig. 1(b), can be fitted equally well with either single or multiple Drude components, the transient THz spectra, shown in Fig. 4, can only be fitted by the two-Drude model. This is the case because the strong constraint imposed by the requirement to simultaneously describe both the conductivity ${\Delta\sigma}_1(\omega)$ and the dielectric function ${\Delta\varepsilon_1(\omega)}$ over a broad temporal and spectral ranges, shown in Figs. 2(a)-2(d), instead of just fitting the lineshape as in the static case. 
Second, an even stronger constraint comes from the {\em intraband}, mid-IR excitation scheme that conserves the fixed number of surface and bulk quasi-particles after the photoexcitation, i.e., carrier densities of both surface and bulk states are kept constant in the fitting, $\Delta n_\text{SS}$=0 and $\Delta n_\text{BS}$=0. Remarkably, by simply varying the surface and bulk scattering rates we are able to consistently reproduce the experimental results, as shown in Fig. 4, and extract the scattering rates (or inverse transport lifetimes that should not be confused with hot carrier cooling times) $\gamma_\text{SS}$ (red line), $\gamma_\text{BS}$ (green line) and their cooling dynamics in Figs. 2(e) and 2(f). 
On the flip side, this excellent agreement corroborates our conclusion that mid-IR photoexcitation creates a unique nonequilibrium state of hot, yet conserved, Dirac fermions and bulk carriers in their respective bands. 
Third, the equilibrium surface and bulk scattering rates, $\gamma_\text{SS}$ and $\gamma_\text{BS}$ extracted before the photoexcitation  are 0.84 THz and 3.18 THz, respectively, which reveal an enhancement ratio, $\gamma_\text{BS}/\gamma_\text{SS }\sim$3.8. Remarkably, this value matches very well with the topological enhancement factor, 4, for protected, helical spin transport of non-interacting fermions in the presence of short-range disorder such as structural defects and surface or interface roughness. In the presence of long-range, Coulomb disorder, this enhancement decreases from 4 to 2 due to increase of screening \cite{TOzturk}. Furthermore, as shown in Figs. 2(e)-2(f), the scattering rates of the surface Dirac and bulk
fermions nearly coincide with the frequency-dependent decay of $\Delta\sigma_1(\omega)$ for 2 meV and 7 meV probe, respectively. This corroborates again that tuning the THz probe frequency under mid-IR pump can be used to disentangle the distinct symmetry-protected transport from the bulk conduction and measure their dynamics without reference to theoretical models. 

\begin{figure}[tbp]
	\includegraphics[width=84mm]{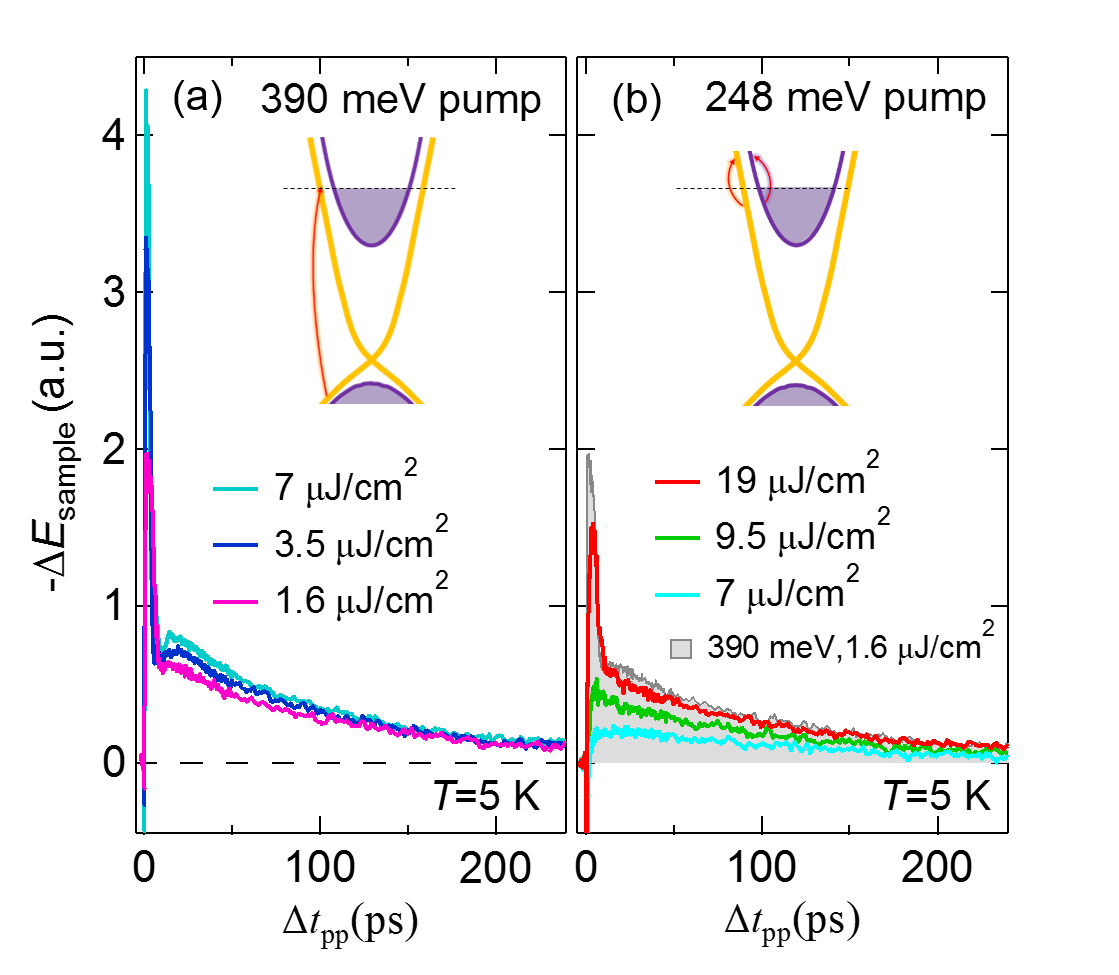}
	\caption{\textbf{Ultrafast THz dynamics under interband and intraband photoexcitations.} Pump-induced THz field changes $\Delta E_\text{sample}$ as a function of pump-probe delay $\Delta t_\text{pp}$ after (a) 390 meV (3.18 $\mu$m) and (b) 248 meV (5 $\mu$m) photoexcitations at various fluences and $T$=5 K. The gate time $t_\text{gate}$ is fixed at the peak of the static THz field shown in Fig. 1(d). One of the THz traces from (a) (pink curve, 1.6 $\mu$J/cm$^{2}$) is also plotted in (b) as a filled gray curve for better comparison. 
	}
	\label{Fig5}
\end{figure}

To highlight the difference between the intraband excitation scheme used in this work and the interband excitation found in previous literature, Figs. 5(a) and 5(b) compare the ultrafast THz dynamics under 390 meV (3.18 $\mu$m) and 248 meV (5 $\mu$m) pumping, respectively. The former resonantly excites interband transitions in our bulk Bi$_{2}$Se$_{3}$ sample (Fig. 1(a)).   
Here, for simplicity, we show the THz field change $\Delta E_\text{sample}$ by fixing the gate time $t_\text{gate}$ at the peak of the static THz field as a function of pump-probe delay $\Delta t_\text{pp}$, similar to previous studies \cite{ RVAguilar,SSim}. Unlike the above-obtained frequency-dependent conductivity ${\Delta\sigma}_1(\omega)$ dynamics that closely follows the pump-induced change of surface/bulk scattering rates, $\Delta E_\text{sample}$ measured at a fixed gate time originates from frequency-integrated responses within the measured spectral range $\sim$2-10 meV. This frequency average leads to more complicated dynamics that may come from all relaxation channels of opposite signs as shown in Fig. 2. Nevertheless, Fig. 5 shows that there is still a distinct difference between the interband (Fig. 5(a)) and intraband (Fig. 5(b)) pump excitation schemes. 
On the one hand, with resonant interband pump excitation at 390 meV (inset, Fig.5(a)), $\Delta E_\text{sample}$ at several pump fluences shows two main decay features: (1) a dominant, fast ``overshoot", which exhibits fluence-dependent amplitude and quickly diminishes within $\sim$4 ps; (2) a slow, 100s of ps component which is fluence independent, i.e., the $\Delta E_\text{sample}$ for all fluences merge and decay together after $\sim$100 ps. 
On the other hand, with intraband excitation at 248 meV (inset, Fig. 5(b)), the fast ``overshoot" component is absent in the THz responses at low fluences $\leqslant$12 $\mu$J/cm$^{2}$, the fluence used for mid-IR pumping in Figs. 1-4. A distinct difference is clearly visible between interband pumping and intraband pumping with low fluence, e.g., the 7 $\mu$J/cm$^{2}$ (cyan) and 9.5  $\mu$J/cm$^{2}$ (green) traces. 
Interestingly, increasing the mid-IR pump fluence to 19 $\mu$J/cm$^{2}$ (red line in Fig. 5(b)) leads to a transition to a decay profile where a fast ``overshoot" component now appears similar to the interband photoexcitation, e.g., the 390 meV excitation scan at 1.6 $\mu$J/cm$^{2}$ (pink line in Fig. 5(a)). For a better comparison, this pink curve is also plotted in Fig. 5(b) as a filled gray curve which resembles the 248 meV excitation scan at 19 $\mu$J/cm$^{2}$, unlike for the lower fluences. 
This reveals the same relaxation mechanism between the two, which can be attributed to the two photon absorption (TPA) present in the high fluence, mid-IR pumping case. This excites interband transitions in our sample in spite of below gap pump photon energy. The dynamics of Bi$_2$Si$_3$ under high photon energy pumping have been extensively studied recently and the few ps relaxation component is mainly from bulk-surface charge transfer that can give the ``overshoot" feature after interband carrier injection \cite{SHEN}. 
The absence of such process in the mid-IR pumping below $\sim$12 $\mu$J/cm$^{2}$ is consistent with our observation of a conserved Dirac fermion transient distribution induced by mid-IR pulses. 

In conclusion, we have provided the first measurement and insights into the {\em intraband}, mid-IR pump-induced THz conductivity. We achieve this by suppressing all interband transitions that mask the processes of main interest which is distinctly different from any prior THz transport measurements so far. This allows us to isolate and measure the intrinsic transport behaviors of Dirac fermions with conserved populations on the surface of doped Bi$_2$Se$_3$. This scheme enables a clear observation of enhanced surface transport, which provides evidence for helical spin structure of surface (Dirac) electrons that suppresses backscattering.
The pump photon energy tuning achieved in this way also allows us to control the surface transport optically in a selective way. Our experiments may evolve into a benchmark characterization method for high frequency topological transport in TI-based device development, and motivate fundamental quantum phase discovery and ultrafast control at the interface between magnetism and electronic topology, e.g., magnetic Weyl semimetals. 

This work was supported by the U.S. Department of
Energy, Office of Basic Energy Science, Division of
Materials Sciences and Engineering (Ames Laboratory is
operated for the U.S. Department of Energy by Iowa State
University under Contract No. DE-AC02-07CH11358).
Sample development was supported by National Science Foundation DMR 1400432.
Terahertz spectroscopy instrument
was supported in part by the M. W. Keck Foundation.

\noindent$^{\ast}$jgwang@iastate.edu; jwang@ameslab.gov

\end{document}